\documentclass[preprint]{aastex}

\usepackage{epsfig}

\slugcomment{Accepted for publication in ApJ Letters}

\shorttitle{Antibias in Clusters}

\begin{document}

\title{Antibias in Clusters: \\
  The Dependence of M/L on Cluster Temperature}
\author{Neta A. Bahcall and Julia M. Comerford}
\affil{Princeton University Observatory, Princeton, NJ 08544}

\begin{abstract}
We show that the observed mass-to-light ratio of galaxy clusters increases with cluster temperature as expected from cosmological simulations. Contrary to previous observational suggestions, we find a mild but robust increase of $M/L$ from poor (T $\sim 1 - 2$ keV) to rich (T $\sim 12$ keV) clusters; over this range, the mean $M/L_V$ increases by a factor of $\sim 2$.  The best fit relation satisfies $M/L_V = (170 \pm 30) T_{keV}^{0.3 \pm 0.1} h$ at $z = 0$, with a large scatter. This trend confirms predictions from cosmological simulations which show that the richest clusters are antibiased, with a higher ratio of mass per unit light than average. The antibias increases with cluster temperature. The effect is caused by the relatively older age of the high-density clusters,  where light has declined more significantly than average since their earlier formation time. Combining the current observations with simulations, we find a global value of $M/L_V = 240 \pm 50 h$, and a corresponding mass density of the universe of $\Omega_m = 0.17 \pm 0.05$. 
\end{abstract}


\clearpage

\section{Introduction}
The mass-to-light ratio of clusters of galaxies has been used for decades in the classical $M/L$ method of estimating the mass density of the universe. In this method, the average ratio of the observed mass to light of clusters of galaxies, which represent systems of relatively large scales ($\sim 1$ Mpc), is assumed to be a fair representation of the mean $M/L$ of the universe. This ratio is then multiplied by the observed luminosity density of the universe to yield the universal mass density. When applied to rich clusters, with average $M/L_B \simeq  300 h$, the total mass density of the universe amounts to $\Omega_m \sim 0.2$ (where $\Omega_m$ is the mass density in units of the critical density; Zwicky 1957; Abell 1965; Ostriker, Peebles, \& Yahil 1974; Bahcall 1977; Faber \& Gallagher 1979; Trimble 1987; Peebles 1993;  Bahcall, Lubin, \& Dorman 1995; Carlberg et al. 1996; Carlberg, Yee, \& Ellingson 1997, and references therein). 

A fundamental assumption in this determination, however, is that the mass-to-light ratio of clusters is a fair representation of the universal value. If $M/L$ of clusters is larger or smaller than the universal mean, the resulting $\Omega_m$ will be an over- or under- estimate, respectively. In general, if mass follows light on large scales ($M/L \simeq$ constant), the galaxy distribution is considered to be unbiased with respect to mass. If mass is distributed more diffusely than light, as is generally believed, then the galaxy distribution is biased (i.e., more clustered than mass), and the above $\Omega_m$ determination would be an underestimate.	

Observations of galaxies, groups, and clusters of galaxies show that the mass-to-light function increases with scale up to a few hundred Kpc \citep{ru70, ro73, os74, ei74, da80, za93, fi00}, but then flattens on larger scales \citep{ba95, fi00}.  A comparison with high resolution cosmological simulations \citep{ba00} reveals an excellent agreement with the observed $M/L$ function: the simulated $M/L$ function increases with scale on small scales ($< 0.5$ Mpc) and flattens to a mean constant value on large scales, as suggested by the data. 

However, the simulations show that while $M/L$ flattens on average on large scales, high overdensity regions, such as rich clusters of galaxies, exhibit higher $M/L$ values than average, while low density regions have lower $M/L$ ratios (at a given scale); high density regions are $antibiased$ in $M/L$, with mass more strongly concentrated than light (a higher $M/L$ ratio) than average. \citet{ba00} show that the $M/L$ antibias exists in simulated clusters of galaxies: higher density clusters (richer, higher mass and temperature clusters) should exhibit larger $M/L$ ratios (especially in the blue and visual luminosity bands) than poorer, lower mass and temperature clusters.  The expected effect is approximately a factor of two increase from poor to rich clusters. It is caused, as seen in the hydrodynamic simulations, by the age of the systems: high density clusters form at earlier times than low density clusters; their luminosities have thus declined more significantly by the present time (if no significant recent star formation occurred), yielding a  larger $M/L$ ratio, on average, for the richer clusters. 

Observationally, however, cluster $M/L$ ratios have been suggested to show no dependence on cluster overdensity (as represented by cluster richness, mass, or temperature) (e.g., Carlberg, Yee, \& Ellingson 1997; Hradecky et al. 2000). In this paper, we investigate the observed $M/L$ dependence on cluster overdensity (temperature) by using a well-studied sample of clusters that range from poor to rich systems. We find that $M/L$ increases, on average, with cluster temperature, in good agreement with the prediction of cosmological simulations. 

\section{Dependence on Cluster Temperature}
We investigate the dependence of $M/L$ on cluster overdensity by using a reliable sample of clusters that ranges from poor to rich systems. The cluster overdensity is represented by the observed temperature of the cluster, T; the overdensity within a given radius is proportional to T: $(\Delta \rho / \rho)_{cl} \propto (M/R^3)_{cl} \propto T$. As revealed by cosmological simulations \citep{ba00}, the $M/L$ ratio of clusters is expected to increase with T by a factor of $\sim 2$ (in $M/L_B$ and $M/L_V$) from poor (T $\sim 1 - 2$ keV) to rich (T $\sim 10$ keV) clusters. 

Current observations of galaxy clusters generally claim no observed dependence of $M/L$ on cluster temperature or other cluster parameters (e.g., Carlberg, Yee, \& Ellingson 1997; Hradecky et al. 2000). However, the expected effect is not large, and the observational uncertainties in $M/L$ are significant. It is important therefore to use a reliable and consistent sample of clusters that spans a wide range of cluster temperatures. 

We select clusters with measured temperatures in the range of $\sim 1$ to $12$ keV. All clusters have measured $M/L$ ratios observed within radii of typically $0.5$ to $1.5 \; h^{-1}$ Mpc. The cluster $M/L$ measurements are based on a few self-consistent subsamples and methods. We analyze different subsamples separately in order to test the reliability of the results. The sample includes clusters with mass determinations from gravitational lensing observations (Fischer \& Tyson 1997; Hoekstra, Franx, \& Kuijken 2000; Squires et al. 1996a, 1996b, 1997; Tyson \& Fischer 1995) and X-ray observations (from a single, carefully executed method; Hradecky et al. 2000). A sample of velocity dispersion virial mass determinations from a single method \citep{ca97} is also shown for comparison, although this method differs somewhat in that it determines cluster virial masses, within the larger (and less certain) virial radii of $\sim 1 - 1.5 \; h^{-1}$ Mpc. We repeat our analysis both with and without this virial subsample and find consistent results. We also repeat the analysis for individual cluster subsamples, e.g., lensing versus X-ray cluster subsamples and high versus low redshift cluster subsamples; all samples yield consistent results. 	

In addition to the cluster sample, we also include the recently observed ensemble average $M/L$ ratio of $50$ groups of galaxies determined from weak gravitational lensing by \citet{ho01}. We convert the mean velocity dispersion for this group sample, $274^{+48}_{-59}$ km/s, to a mean temperature using the observed velocity-temperature relation for clusters and groups: $V_r = (332 \pm 52) T_{keV}^{0.6}$ km/s \citep{lu93}. This yields a mean group temperature of T $= 0.73$ keV, consistent with observations of typical groups ($\sim 1$ keV). The total sample we use, $21$ systems, is listed in Table 1. Also listed are the redshifts, the observed temperatures and $M/L$ ratios, the luminosity band of the observations (mostly V band), the relevant radius $R$ for the observed parameters, and the relevant references. A Hubble constant of $H_0 = 100 h$ km/s/Mpc is used throughout. 	
      
Two corrections are applied to the sample to bring all measurements to the same consistent system. First, the sample contains clusters at redshifts from $z \sim 0.02$ to $0.8$. Since cluster luminosities evolve with redshift, we correct all luminosities to $z = 0$. Given the observed evolution in $L_V$ \citep{ke97}, $L_V \sim 10^{(0.3z \pm 10\%)}$, all $M/L_V$ values were corrected upward by a factor of $10^{(0.3z \pm 10\%)}$ to yield $M/L_V (z = 0)$. Similarly, $M/L_B$ values were corrected by a factor of $10^{(0.4z \pm 10\%)}$ \citep{va98} and $M/L_r$ values were corrected by a factor of $10^{(0.15z \pm 10\%)}$ \citep{ca96}.  

Second, four of the $21$ clusters are observed in the blue band luminosity, $L_B$, and six are observed in the $r$ band luminosity, $L_r$ (all others are observed in $L_V$). We convert $L_B$ to $L_V$ at $z = 0$ using $L_V \simeq 1.15 \; L_B$, based on the typical cluster color of $B - V \simeq 0.8$ (and solar $(B - V)_\odot = 0.65$). Similarly, we convert $L_r$ to $L_V$ using $L_V \simeq 0.935 \; L_r$, following $(B - r)_{cl} \simeq 0.915$ and $(B - r)_\odot = 0.692$ \citep{jo95}.  These corrections have no significant impact on the final results. In addition, all cluster values (and the evolution correction) are corrected to the same cosmology of $q_0 = 0.15$ (following van Dokkum et al. 1998). The resulting $M/L_V$ values at $z = 0$ are listed in Table 1. The error bars represent the convolved error bars of the observations and the evolutionary correction.

The sample includes cluster $M/L$ ratios measured within radii of  $\sim 0.5$ to $1.5 \; h^{-1}$ Mpc. The results are not significantly affected by the different radii; we test this by comparing a subsample of clusters measured at a fixed radius ($0.5 \; h^{-1}$ Mpc) and find consistent results. The cluster $M/L$ ratio is nearly constant over this range of radii within a given cluster. 

The observed $M/L_V (z = 0 )$ ratios are plotted as a function of cluster temperature for all clusters in Figure 1. While the scatter is large, a clear correlation of increasing $M/L_V$ with T is apparent. The best fit linear and power-law relations are:

\begin{equation}
%\begin{array}{ll}
\left(\frac{M}{L_V}\right)_{z=0} = 142 \pm 32 + (23 \pm 5) \; T_{keV} \; h \\
= 142 \pm 32 [1 + (0.16 \pm 0.05) \; T_{keV}] \; h    
%\end{array}     
\end{equation}

\begin{equation}
\left(\frac{M}{L_V}\right)_{z=0} = (173 \pm 29) \; T_{keV}^{0.30 \pm 0.08} \; h               
\end{equation}

These best fit relations are shown in Figure 1. Also shown in the figure is the relation predicted by the cosmological simulations \citep{ba00}. The agreement between data and simulations is excellent; both show an increase in $M/L_V$ by a factor of $\sim 2$ on average, from $\sim 200 h$ to $450 h$, as T increases from $\sim 1$ to $12$ keV. While the simulations $M/L (T)$ trend is insensitive to $\Omega_m$, the normalization, or absolute value of $M/L$ at a given T, is proportional to $\Omega_m$. The simulation results plotted in Figure 1, which best fit the data, correspond to $\Omega_m=0.17$ (using the observed $(M/L_V)_{critical} \simeq 1400h$ as the simulation normalization; see Bahcall et al. 2000). 

Figure 2 represents the same data as Figure 1, but with the data binned in several temperature bins ($3$ keV each). The average $M/L_V$ for each bin is shown. The best fit relations, also plotted, are consistent with those obtained for the full sample (Figure 1, eq. [1], eq. [2]). The best fit results and their $\chi^2$ are summarized in Table 2.

To test the sensitivity of the results to the cluster selection method and to the radius used, we repeat the analysis for two subsamples of the full sample: 1) we omit the clusters with dynamical determination of virial masses (to $R_v \sim 1 - 1.5 \; h^{-1}$ Mpc) based on velocity dispersion data \citep{ca97}; and 2 ) we use only clusters measured within a fixed radius of $0.5 \; h^{-1}$ Mpc.  The best fit results obtained for each of these subsamples are consistent with the results of the full sample; they are summarized in Table 2. The results for the $R = 0.5 \; h^{-1}$ Mpc subsample are shown in Figure 3.

We further test the results by analyzing other subsamples of clusters: clusters with gravitational lensing masses versus clusters with X-ray mass determinations, as well as clusters at low redshift ($z < 0.1$) versus clusters at intermediate redshift ($z \sim 0.15$ to $0.33$). All subsamples yield consistent results (within $1 \sigma$). These tests reinforce the robustness of the results.

The above results are in excellent agreement with the prediction of cosmological simulations, showing an increase in cluster $M/L_V$ with cluster temperature (or overdensity). 

\section{Summary and Conclusions}
We show that the mass-to-light ratio of clusters of galaxies increases, on average, with cluster temperature. Contrary to previous observational suggestions of a constant $M/L$ ratio for clusters  -- independent of temperature, velocity dispersion, or other cluster parameters -- we show a consistent trend of increasing $M/L_V$ with temperature from groups and poor clusters (at T $\sim 1$ keV) to the richest clusters (T $\sim 12$ keV); over this range, the mean $M/L_V$ ratio increases by a factor of $\sim 2$. The best fit relations are given by equation (1) and equation (2) (and Table 2, Figures 1 - 3). 

This trend is in excellent agreement with the prediction of cosmological simulations \citep{ba00} (see Figures 1 - 3 for comparisons between data and simulations). While the scatter is large, the underlying trend is clear.

The observed correlation is important for several reasons.

1) It confirms the predictions originally made from cosmological simulations \citep{ba00} of increasing cluster $M/L$ ratio with cluster overdensity (temperature). As seen in the simulations, this trend is caused by the ``age effect'': higher density systems formed earlier and their luminosities, especially in the blue and visual bands, have decreased more significantly by the present time relative to the younger, lower density systems.  This results in larger $M/L$ ratios for the older, higher temperature clusters. 

2) The simulated results, confirmed by the current observations, show that rich clusters are antibiased in their $M/L$ ratios -- that is, rich clusters have higher $M/L$ ratios than average.  This implies that mass is more concentrated than light, on average, in rich clusters (Bahcall et al. 2000; see also Jing, Mo, \& Borner 1998). This antibias increases with cluster richness (temperature).

3) Antibias in the cluster $M/L$ ratio directly affects the measurement of the mass density of the universe, $\Omega_m$. Classically, $\Omega_m$ determinations use the observed $M/L$ ratio of the richest clusters (which are easiest to observe) and assume it to be representative of the global value. The observed antibias, however, shows that this method overestimates $\Omega_m$, since rich clusters have larger $M/L$ ratios than average. Comparing the observed dependence of $M/L$ on T with results from the cosmological simulations \citep{ba00}, shown in Figures 1 - 3 [using $(M/L_V)_{critical}=1400h \pm 20\%$ for the simulation normalization (see Bahcall et al. 2000)], we find a global mass density parameter of $\Omega_m = 0.17 \pm 0.05$. The mean representative mass-to-light ratio for the universe is $M/L_V = 240 \pm 50 h$, comparable to that exhibited by groups and poor clusters.

\clearpage

\begin{deluxetable}{cccccccc}
\tabletypesize{\scriptsize}
\tablecaption{Cluster parameters. \label{tbl-1}}
\tablewidth{0pt}
\tablehead{
\colhead{Object} & 
\colhead{$z$} & 
\colhead{T}   &
\colhead{$M/L$} &
\colhead{Band}  & \colhead{$R$} & \colhead{$M/L_V$} &
\colhead{Ref.}  \\
\colhead{ } & \colhead{ } & \colhead{(keV)} & \colhead{($h \; M_\odot/L_\odot$)} &
\colhead{ } & \colhead{($h^{-1}$ Mpc)} & \colhead{($z=0$)} & \colhead{ }
}
\startdata
A262 & 0.016 & $2.3 \pm 0.2$ & $180 \pm 64$ & V & 0.5 & $180 \pm 64$ & 7 \\
A426 & 0.018 & $6.2 \pm 0.4$ & $154 \pm 68$ & V & 0.5 & $154 \pm 68$ & 7 \\
A478 & 0.09 & $8.4^{+0.8}_{-1.4}$ & $276^{+124}_{-218}$ & V & 0.5 & $288^{+129}_{-227}$ & 7 \\
A1795 & 0.062 & $7.8 \pm 1.0$ & $468^{+148}_{-140}$ & V & 0.5 & $479^{+151}_{-143}$ & 7 \\ 
A2052 & 0.035 & $2.8 \pm 0.2$ & $198 \pm 54$ & V & 0.5 & $201 \pm 55$ & 7 \\
A2063 & 0.035 & $2.3 \pm 0.2$ & $190 \pm 46$ & V & 0.5 & $193 \pm 47$ & 7 \\
A2199 & 0.030 & $4.8 \pm 0.1$ & $294 \pm 110$ & V & 0.5 & $297 \pm 111$ & 7 \\
MKW4s & 0.028 & $1.8 \pm 0.3$ & $202 \pm 86$ & V & 0.5 & $204 \pm 87$ & 7 \\
A2163 & 0.20 & $11.0 \pm 0.6$ & $300 \pm 100$ & V & 0.5 & $331 \pm 110$ & 8, 12 \\ 
MS1054 & 0.83 & $12.3^{+3.1}_{-2.2}$ & $248 \pm 34$ & B & 0.5 & $464 \pm 74$ & 2, 5 \\
A2218 & 0.18 & $7.48^{+0.53}_{-0.41}$ & $440 \pm 80$ & B & 0.4 & $478 \pm 87$ & 4, 10 \\
A1689 & 0.18 & $9.02^{+0.40}_{-0.30}$ & $400 \pm 60$ & V & 1 & $435 \pm 65$ & 4, 13 \\
A2390 & 0.23 & $8.90^{+0.97}_{-0.77}$ & $320 \pm 90$ & V & 1 & $360 \pm 101$ & 4, 11 \\
RXJ1347 & 0.45 & $11.37^{+1.10}_{-0.92}$ & $200 \pm 50$ & B & 1 & $264 \pm 67$ & 4, 3 \\
MS0016 & 0.55 & $8.0 \pm 1.0$ & $213 \pm 74$ & r & 1.32 & $275 \pm 96$ & 4, 1\\
MS1358 & 0.33 & $6.50^{+0.68}_{-0.64}$ & $188 \pm 40$ & r & 1.15 & $225 \pm 48$ & 4, 1 \\
MS0440 & 0.20 & $5.3^{+1.3}_{-0.9}$ & $314 \pm 105$ & r & 0.87 & $360 \pm 120$ & 9, 1 \\
MS0451 & 0.54 & $10.17^{+1.55}_{-1.26}$ & $314 \pm 105$ & r & 1.45 & $405 \pm 136$ & 9, 1 \\
MS0839 & 0.19 & $4.19^{+0.36}_{-0.33}$ & $317 \pm 132$ & r & 1.12 & $362 \pm 151$ & 9, 1 \\
MS1008 & 0.31 & $7.29^{+2.45}_{-1.52}$ & $247 \pm 79$ & r & 1.36 & $294 \pm 94$ & 9, 1 \\
50 groups & 0.33 & $0.73^{+0.07}_{-0.08}$ & $191 \pm 83$ & B & 0.5 to 1 & $225 \pm 98$ & 6 \\
\enddata

\tablerefs{
(1) Carlberg, Yee, \& Ellingson 1997; (2) Donahue et al. 1998; (3) Fischer \& Tyson 1997; (4) Hjorth, Oukbir, \& van Kampen 1998; (5) Hoekstra, Franx, \& Kuijken 2000; (6) Hoekstra et al. 2001; (7) Hradecky et al. 2000; (8) Irwin \& Bregman 2000; (9) Mushotzky \& Scharf 1997; (10) Squires et al. 1996a; (11) Squires et al. 1996b; (12) Squires et al. 1997; (13) Tyson \& Fischer 1995.}

\end{deluxetable}

\clearpage

\begin{deluxetable}{ccccc}
\tabletypesize{\scriptsize}
\tablecaption{Best-fit $M/L_V$ - T relations. \label{tbl-2}}
\tablewidth{0pt}
\tablehead{
\colhead{Sample} & 
\colhead{$M/L_V = A \; T_{keV} + B$} &
\colhead{$\chi^2/dof$} &
\colhead{$M/L_V = C T_{keV}^\alpha$}  & 
\colhead{$\chi^2/dof$} 
}
\startdata

Full sample (21) & $(23 \pm 5)T+(142 \pm 32)$ & $0.8$ & $(173 \pm 29)T^{0.30 \pm 0.08}$ & $1.0$ \\
Subsample $1\tablenotemark{*}$ (15) & $(21 \pm 6)T+(153 \pm 39)$ & $1.2$ & $(171 \pm 33)T^{0.30 \pm 0.10}$ & $1.3$ \\
Subsample $2\tablenotemark{*}$ (11) & $(26 \pm 7)T+(129 \pm 40)$ & $1.0$ & $(133 \pm 35)T^{0.44 \pm 0.14}$ & $1.1$ \\
\enddata

\tablenotetext{*}{Subsample 1: excludes clusters with dynamical determination of virial masses} 
\tablenotetext{*}{Subsample 2: clusters observed at a fixed radius ($R = 0.5 \; h^{-1}$ Mpc)}

\end{deluxetable}

\clearpage

\begin{figure}
\plotone{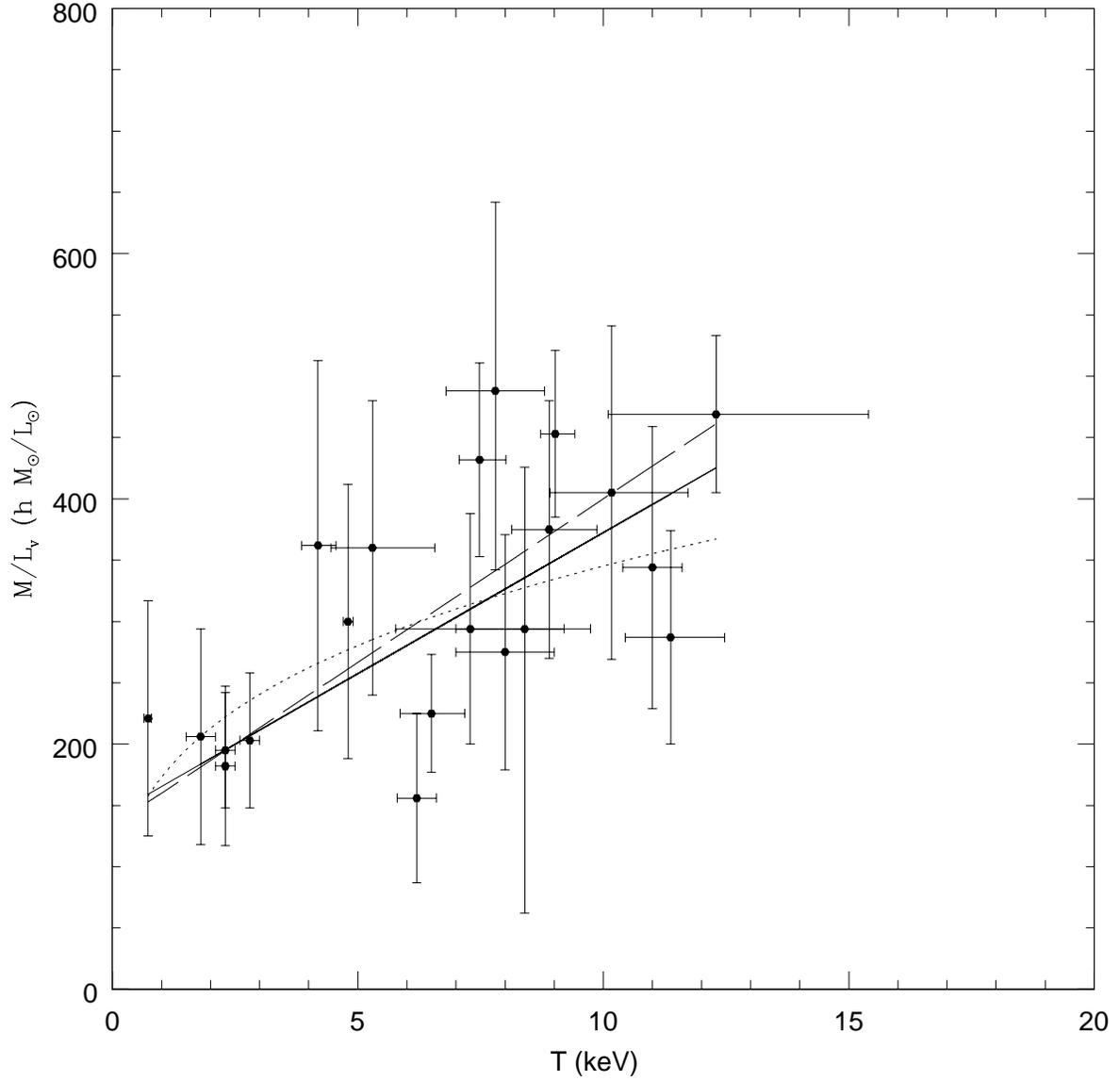}
\figcaption{Observed $M/L_V$ (corrected to $z = 0$) vs. T for clusters of galaxies.  The best fit linear and power law relations (eq. [1] and eq. [2]) are presented by the solid and dotted lines, respectively.  The prediction from simulations is shown by the dashed line; its normalization corresponds to $\Omega_m = 0.17$. \label{fig1}}
\end{figure}

\clearpage

\begin{figure}
\plotone{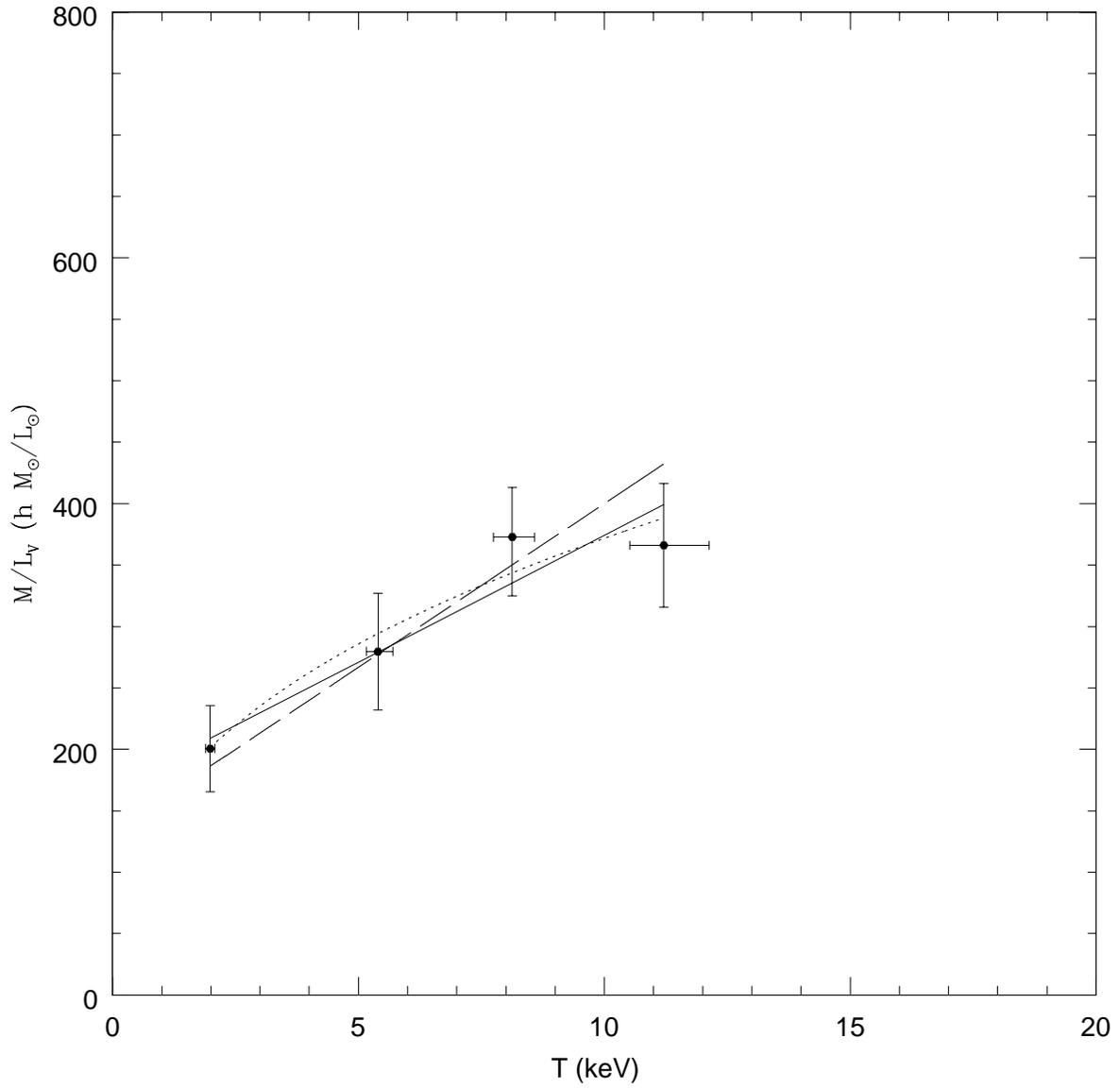}
\figcaption{Same as Figure 1 but for the data binned in 3 keV temperature bins. \label{fig2}}
\end{figure}

\clearpage

\begin{figure}
\plotone{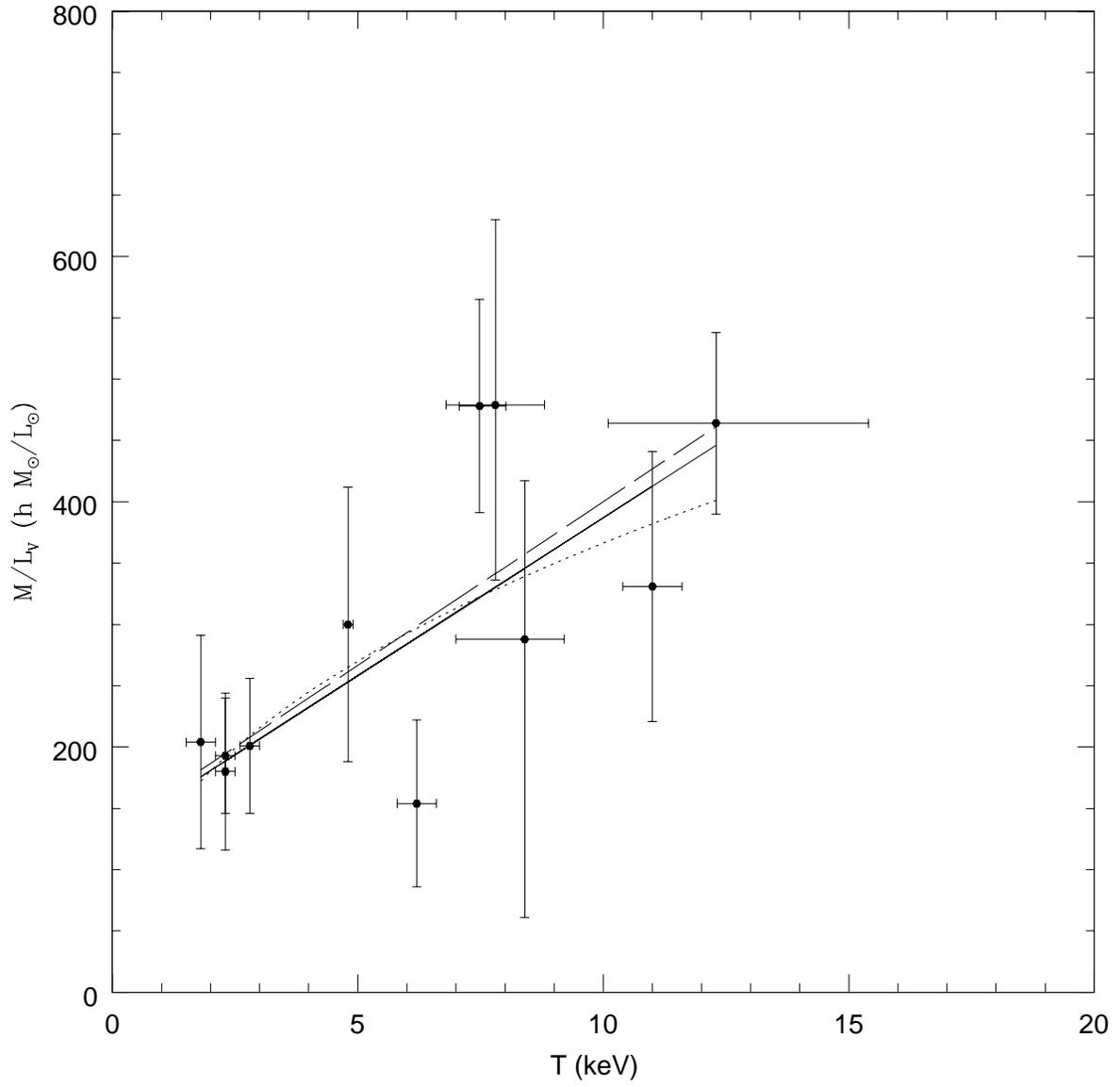}
\figcaption{$M/L_V$ (corrected to $z = 0$) vs. T for clusters at $R = 0.5 \; h^{-1}$ Mpc. \label{fig3}}
\end{figure}

\end{document}